\documentstyle{pasj06}
\begin{document}

\SetRunningHead{A.Imada et al.}{}

\title{Discovery of superhumps during a normal outburst of SU Ursae
Majoris}

\author{Akira \textsc{Imada}$^1$, Hideyuki \textsc{Izumiura}$^1$,
Daisuke \textsc{Kuroda}$^1$, Kenshi \textsc{Yanagisawa}$^1$, \\
Nobuyuki \textsc{Kawai}$^2$, Toshihiro \textsc{Omodaka}$^3$, and Ryo \textsc{Miyanoshita}$^3$}

\affil{$^1$Okayama Astrophysical Observatory, National Astronomical
Observatory of Japan, Asakuchi,\\ Okayama 719-0232}
\affil{$^2$Department of Physics, Tokyo Institute of Technology,
Ookayama 2-12-1, Meguro-ku, Tokyo 152-8551}
\affil{$^3$Faculty of Science, Kagoshima University, 1-21-30 Korimoto,
Kagoshima, Kagoshima 890-0065}


\maketitle

\begin{abstract}

We report on time-resolved photometry during a 2012 January normal
 outburst of SU UMa. The light curve shows hump-like modulations with
 a period of 0.07903(11) d, which coincides with the known superhump
 period of SU UMa during superoutbursts. We interpret this as superhump,
 based on the observed periodicity, profiles of the averaged light
 curve, and the $g'-I_{\rm c}$ variation during the normal
 outburst. This is the first case that superhumps are detected during an
 isolated normal outburst of SU UMa-type dwarf novae. The present result
 strongly suggests that the radius of the accretion disk already reaches
 the 3:1 resonance even in the midst of the supercycle. 

\end{abstract}

\section{Introduction}

SU UMa-type dwarf novae, a subclass of dwarf novae, exhibit two types
of outbursts: normal outburst and superoutburst (for a review, see
\cite{war95book}; \cite{osa96review}). During superoutburst,
hump-like modulations called superhumps are visible. Basically, the
light source of superhumps is understood as phase-dependent tidal
dissipation in an eccentric accretion disk (\cite{whi88tidal};
\cite{hir90SHexcess}). The general consensus is that an eccentricity of
the accretion disk is excited by a 3:1 orbital resonance
(\cite{osa89suuma}). According to the
original thermal-tidal instability model (TTI model, \cite{osa89suuma}),
the radius of the accretion disk monotonically increases with each
normal outburst. When the disk finally reaches the 3:1 resonance radius,
the accretion disk is tidally deformed and triggers a superoutburst. The
TTI model reproduces well basic behavior of SU UMa-type dwarf
novae. However, a reform of the TTI model may be required, particularly in
systems with unusual recurrence times of superoutbursts
(\cite{hel01eruma}; \cite{pat02wzsge}; \cite{osa03DNoutburst}).

Over the past few years, research activity concerning SU UMa-type dwarf
novae has been significantly improved. One of the most important research
is that unprecedented photometric surveys during superoutbursts have
been carried out by T. Kato and his colleagues (\cite{pdot}; \cite{pdot2};
\cite{pdot3}). They collected all of available data and
analyzed the light curves during superoutbursts, from which they have
established the basic picture of the superhump period changes (see
figure 3 of \citet{pdot}). Another important research to be noted is
that the $Kepler$ satellite has provided us with unprecedentedly precise
light curves at the one-minite cadence (\cite{kepler1};
\cite{kepler2}). This allows us to investigate detailed light curves
that cannot be achieved under ground-based observations
(\cite{2010ApJ...717L.113S}; \cite{2010ApJ...725.1393C};
\cite{2011ApJ...741..105W}; \cite{pdot3};
\cite{2012ApJ...747..117C}). Although these surveys improve our
understanding of SU UMa-type dwarf novae, the diversity of the
observations claims further modification of the TTI model.

In order to decipher and understand the observed diversity of SU
UMa-type dwarf novae, we started a new approach: simultaneous multicolor
photometry of dwarf novae not only during outburst but also during
quiescence. As a first step, we performed multicolor photometry of SU
UMa itself from 2011 December to 2012 February. SU UMa is a prototype of
SU UMa-type dwarf novae ranging $V$=11.3-15.7 \citep{RKcat08} and its
orbital period is determined to be $P_{\rm orb}$=0.07635 d
\citep{tho86suuma}. However, anomalous behavior with short and long time
scales were reported in the previous studies (\cite{ros00suuma};
\cite{kat02suuma}). In this letter, we report on detection of superhumps
during a 2012 January normal outburst. This is the first recorded
superhumps that emerged in the middle of a supercycle. Results of the
whole observations will be discussed in a forthcoming paper.

\section{Observation and Result}

\begin{figure*}
\begin{center}
\FigureFile(160mm,80mm){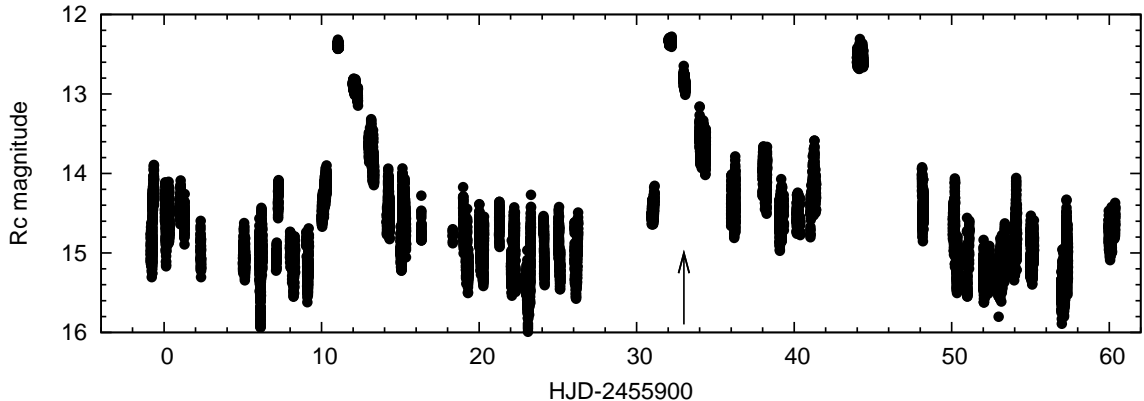}
\end{center}
\caption{$R_{\rm c}$ band light curve of SU UMa. The abscissa and
 ordinate denote HJD$-$2455900 and $R_{\rm c}$ magnitude,
 respectively. The normal outburst in which superhumps are detected is
 marked with an arrow. Note that this normal outburst is held between
 two normal outbursts.}
\label{lc}
\end{figure*}

Time-resolved CCD photometry were performed from 2011 December 1 to 2012
February 20 at Okayama Astrophysical Observatory using the 50-cm MITSuME
telescope, which is able to obtain $g'$, $R_{\rm c}$, and $I_{\rm c}$
bands simultaneously \citep{3me}. In this letter, we extracted data from
2012 January 4 to 7, during which SU UMa experienced a normal
outburst. The exposure time was 30 s with a read-out time as short as 1
s. The data were processed under the standard manner using
IRAF/daophot\footnote{IRAF (Image Reduction and Analysis Facility) is
distributed by the National Optical Astronomy Observatories, which is
operated by the Association of Universities for Research in Astronomy,
Inc., under cooperative agreement with National Science
Foundation.}. After removing bad images, we acquired
available 1577 images for $g'$ band, 1588 images for $R_{\rm c}$ band,
and 1587 images for $I_{\rm c}$ band, respectively. Differential
photometry were carried out using Tycho-2 4126-00036-1 (RA:
08:12:45.104, Dec: +62:26:17.57), whose constancy was checked by nearby
stars in the same image. Heliocentric correction was made before the
following analyses.

\begin{figure}
\begin{center}
\FigureFile(80mm,80mm){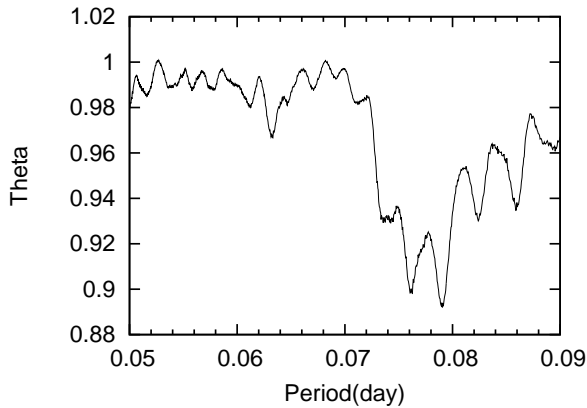}
\end{center}
\caption{PDM analysis during HJD 2455932-34, corresponding to the
 declining stage of the normal outburst. A weak signal can be seen at
 $P$=0.07903(11) d, almost identical to the mean superhump period during
 the 1989 April superoutburst of SU UMa. Also seen is the periodicity at
 $P$=0.07616(11) d, slightly shorter than the orbital period of SU
 UMa.}
\label{pdm}
\end{figure}

Figure \ref{lc} shows $R_{\rm c}$ band light curve of SU UMa between
2011 December 3 and 2012 February 2. On 2012 January 4 (HJD 2455931),
the magnitude monotonically brightened at a rate of -1.41(4) mag/d,
indicating the initiation of the outburst. The magnitude reached
$R_{\rm c}$ ${\sim}$ 12.3 on 2012 January 5 (HJD 2455932), after which
the light curve decayed at a rate of 0.64(1) mag/d. After the end of the
normal outburst, surprisingly, SU UMa entered an anomalous state in
which the magnitude was ${\sim}$ 0.5 mag brighter than that in usual
quiescence. This $bright$ $quiescence$ lasted until the next normal
outburst. More details are beyond the scope of this letter and will be
discussed in a forthcoming paper. Taking this observation into
consideration, we infer that the normal outburst ended until 2012
January 8 (HJD 2455935).

We performed the phase dispersion minimization method (PDM,
\cite{ste78pdm}) for estimation of periods during the normal
outburst. After removing the declining trends, we combined light curves
on 2012 January 5, 6, and 7. The resultant theta diagram on $R_{\rm c}$
band is displayed in figure \ref{pdm}. As can be seen in this figure,
two strong signals coincide with $P$=0.07616(11) d and $P$=0.07903(11) d,
respectively. The former period is very close to the orbital period of
the system but slightly shorter. The latter period, on the other hand,
is in excellent agreement with the mean superhump period during
the 1989 April superoutburst of SU UMa reported by \citet{uda90suuma}.

In order to clarify the nature of this periodicity, we obtained
phase-averaged $R_{\rm c}$ band light curve and $g'-I_{\rm c}$ color
folded with 0.07903 d, which are given in figure \ref{shvar}. Although
the data contain a secondary peak around phase 0.4, a rapid rise and
slow decline, characteristic of superhumps are visible. Furthermore,
this profile bears significant resemblance to that obtained in the
previous study \citep{kat02suuma}. As for $g'-I_{\rm c}$ color, the
bluest peak is prior to the maximum timing of the $R_{\rm c}$
light curve by ${\phi}{\sim}$0.2. It should be noted again that such a
phase discordance also occurred during the 2007 superoutbursts of
V455 And \citep{mat09v455and}.\footnote{See also \citet{uem08j1021} in
which phase discordance between $V$ and $J$ was reported during the 2006
superoutburst of SDSS J102146.44+234926.3.} We also folded the
$R_{\rm c}$ light curve and $g'-I_{\rm c}$ with 0.07616 d, which are
given in figure 4. In this figure, a significant difference compared
with figure 3 is that the peak $R_{\rm c}$ magnitudes coincide with the
bluest peaks in $g'-I_{\rm c}$.

\begin{figure}
\begin{center}
\FigureFile(80mm,80mm){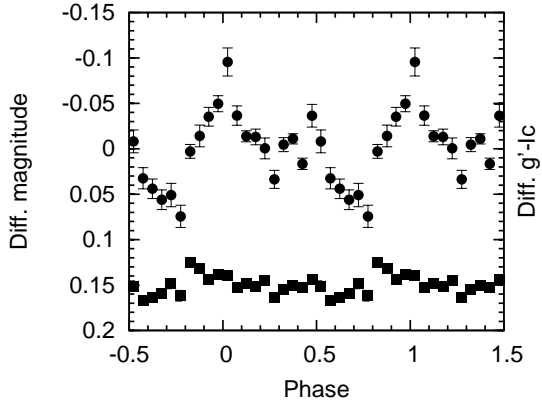}
\end{center}
\caption{Phase averaged $R_{\rm c}$ (filled circle) and $g'-Ic$ color
 (filled square) after folding with
 $P$=0.07903 d. Although $R_{\rm c}$ light curve shows the secondary peak, a
 rapid rise and slow decline, characteristic of superhumps are
 visible. Note that the bluest peak in $g'-I_{\rm c}$ is prior to
 $R_{\rm c}$ by phase ${\sim}$ 0.2. Datapoints are vertically shifted
 for display purpose.}
\label{shvar}
\end{figure}

\begin{figure}
\begin{center}
\FigureFile(80mm,80mm){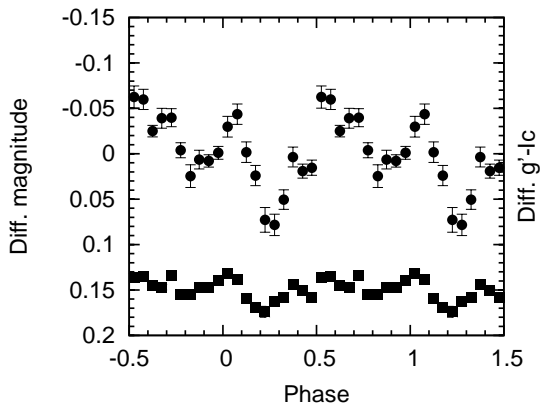}
\end{center}
\caption{Same as figure 3 but folding with $P$=0.07916 d. Double-peaked
modulations, remiscent of orbital humps, are visible. Note that the
bluest peak in $g'-I_{\rm c}$ is in accordance with the magnitude peak
in $R_{\rm c}$. Datapoints are vertically shifted for display purpose.}
\label{shvar}
\end{figure}

\section{Discussion}

In the previous studies, many authors have reported the presence of
(late) superhumps after the end of the superoutburst (\cite{pat02wzsge};
\cite{pat98egcnc}; \cite{uem02j2329}; \cite{kat04egcnc};
\cite{kat08lsh}; \cite{pdot}). This phenomenon can be understood if the
accretion disk persist on eccentricity after the end of the
superoutburst. Recently, \citet{pdot3} detected superhumps during a
normal outburst of V1504 Cyg. After this normal outburst, V1504 Cyg
returned to quiescence and the subsequent outburst was erupted as
superoutburst. In this case, full development of superhumps may have
been prevented despite the radius of the accretion exceeding the 3:1
resonance. In any case, superhumps are observed in the vicinity of a
superoutburst.

According to the AAVSO light curve generator, SU UMa experienced
superoutbursts on 2011 July and 2012 May. This means that the 2012
January normal outburst is independent of the superoutbursts. As for
magnitudes and color behavior, \citet{mat09v455and} suggest that
superhump phase discordance between them is associated with the heating
or cooling process in the accretion disk. In combination with the
observed periodicity, profile of the light curve, and behavior of color
index in $g'-I_{\rm c}$, we can conclude that this is the first example
that superhumps are observed during an $isolated$ normal outburst of SU
UMa-type dwarf novae. Our present result further indicate that the
radius of the accretion disk exceeds the 3:1 resonance radius even in the
middle of the supercycle. Recently, \citet{2012ApJ...747..117C} studied
Kepler data of V344 Lyr, in which the radius at which the
thermal instability sets in ($r_{\rm trig}$) may be the largest roughly
in the midst of a supercycle. Although $r_{\rm trig}$ may not be
necessarily linked with the radius of the accretion disk, this result
provides the potential possibility that the accretion disk exceeds the
3:1 resonance radius even in the midst of quiescence. In order to test
our hypothesis, quiescent spectroscopy should be performed, which
enables us to measure the radius of the accretion disk.

Finally, we briefly discuss figure 4. As already described above, the
most important point is that the peak magnitude coincides with color
index in $g'-I_{\rm c}$. This implies that the main light source is
influenced by a geometric effect of the accretion disk rather than the
disk process itself. If this is the case, then negative superhumps in ER
UMa stars, possibly originated from a tilted disk, show phase accordance
between magnitude and color \citep{oht12eruma}. This should be
clarified in future observations.

\vskip 5mm

We express our gratitude to Daisaku Nogami for his constructive comments
on the manuscript of the letter. We acknowledge with thanks the variable
star observations from the AAVSO International Database contributed by
observers worldwide and used in this research. A.I. and H.I. are
supported by Grant-In-Aid for Scientific Research (A) 23244038 from
Japan Society for Promotion of the Science (JSPS). This work is partly
supported by Optical \& Near-infrared Astronomy Inter-University
Cooperation Program, supported by the MEXT of Japan.


\vskip 20mm

\end{document}